\documentclass[11pt,aps]{revtex4-1}

\usepackage{graphicx}
\usepackage{natbib}
\usepackage{wrapfig}

\newcommand{\be}{\begin{equation}}
\newcommand{\ee}{\end{equation}}
\newcommand{\nn}{\mbox{} \nonumber \\ \mbox{}}
\newcommand{\ba}{\begin{eqnarray}}
\newcommand{\ea}{\end{eqnarray}}
\newcommand{\om}{\omega}

\newcommand{\E}{{\bf E}}
\newcommand{\B}{{\bf B}}

\newcommand{\Bf}{{magnetic field}}

\newcommand{\NS}{neutron star}
\newcommand{\ms}{magnetosphere}
\newcommand{\NSs}{{neutron stars}}

\newcommand{\BH}{{black hole}}
\newcommand{\EM}{{electromagnetic}}
\newcommand{\Sch}{Schwarzschild}

\def\half{\frac{1}{2}}
\def\part{\partial}

\def\a{\alpha}
\def\b{\beta}

\def\eps{\epsilon}

\def\kap{\kappa}
\def\ome{\omega}

\def\Lam{\Lambda}
\def\gam{\gamma}

\def\cL{{\cal L}}

\def\cE{{\cal E}}

\def\x{{\bf x}}

\def\ta{\tilde{a}}

\def\hL{\hat{L}}
\def\Azero{A^{(0)}}
\def\Aone{A^{(1)}}
\def\Fzero{F^{(0)}}
\def\cE{{\cal E}_0}
\def\cB{{\cal B}_0}

\newcommand\eg{\it{e.g.,}}

\newcommand\lo{\mathrel{\raise.3ex\hbox{$<$}\mkern-14mu\lower0.6ex\hbox{$\sim$}}}
\newcommand\go{\mathrel{\raise.3ex\hbox{$>$}\mkern-14mu\lower0.6ex\hbox{$\sim$}}}

\def\be{\begin{equation}}
\def\ee{\end{equation}}
\def\half{\frac{1}{2}}
\def\part{\partial}

\def\a{\alpha}
\def\b{\beta}

\def\eps{\epsilon}

\def\kap{\kappa}
\def\ome{\omega}

\def\Lam{\Lambda}
\def\gam{\gamma}

\def\cL{{\cal L}}

\usepackage{amsmath}
\usepackage[normalem]{ulem}
\usepackage{color}

\begin{document}

\title{Cherenkov emission by a fast-moving  uncharged  \Sch\ \BH}

\author{Sergei Khlebnikov and Maxim Lyutikov\\
 Department of Physics and Astronomy, Purdue University, \\  525 Northwestern Avenue, West Lafayette, IN 47907-2036, USA}

\begin{abstract}
We demonstrate that, in the presence of an external \Bf,  an uncharged  classical \Sch\  \BH\  moving  superluminally in a dielectric with permittivity $\epsilon > 1$ 
produces  Cherenkov emission. 
This is a new  physical effect: classical (non-quantum)  emission of \EM\ waves by  a  completely  charge-neutral ``particle.''
The governing equations (involving general relativity, electromagnetism,  and the physics of  continuous media)  have no external \EM\ source---it is the 
distortion of the initial  \EM\ fields by the gravity of the \BH\ that  plays the role of a superluminally moving  source.  The effect relies on nonzero values of both the magnetic field and the gravitational radius, as well as on the usual Cherenkov condition on the velocity, $v/c > 1/\sqrt{\eps}$.
 Unlike Cherenkov emission by a point charge, the effective source in this case is spatially distributed,   with emission generated along the single Cherenkov emission cone.
 The  emitted spectrum is red-dominated, with  power $\propto dk_z /|k_z|$ for wave numbers  $|k_z| \leq 1/R_G$, where $R_G$ is the \Sch\ radius. 
 We comment on possible observability of this process during \BH --\NS\ mergers.
\end{abstract}

\maketitle

\section{Introduction}
There is a number of classical (non-quantum) effects at the intersection of general relativity (GR), electromagnetism  (EM) and physics of continuous media. Arguably,  the most famous one is the  Blandford-Znajek (BZ) effect  \citep{1977MNRAS.179..433B}, whereby a Kerr \BH\ (BH) immersed in plasma in \Bf\ produces jets by extracting the rotational energy of the BH. Other related phenomena  include  linear motion of a \BH\ through \Bf\ \citep{2010Sci...329..927P,2011PhRvD..83f4001L} producing \EM\ outflows, and the violation of no-hair theorem due to plasma conductivity \cite{2011PhRvD..83l4035L,2011PhRvD..84h4019L,2021PhRvL.127e5101B}.

Various  regimes of GR-EM-plasma interaction  are  becoming important for Multi-messenger astronomy \citep[mergers of double \NSs,  NS-BH, and double BHs,][]{2016PhRvX...6d1015A,2017ApJ...848L..12A}. Particularly interesting is possible detection of precursor emission to compacts' mergers \citep{2024arXiv240216504L}, as  interactions of merging NSs is expected to produce a range of possibly observable  phenomena
\cite{2001MNRAS.322..695H,2019MNRAS.483.2766L,2020ApJ...893L...6M,2023PhRvE.107b5205L}.

In this paper we discuss a novel  effect at the intersection of gravity, electromagnetism and physics of continuous media:  Cherenkov emission by a  {\it classical  uncharged} \Sch\ \BH\ moving through a dielectric in \Bf. This  possibly can produce  an \EM\ precursor to  LIGO events.

The effect under investigation is a fundamentally  novel one. Conventionally, Cherenkov emission is the  emission of  \EM\ waves by a charged particle moving superluminally in a medium with refractive index larger than unity.
Even in the case of Cherenkov emission by a charge-neutral  electric dipole, there are two physical charges, separated in space, that eventually produce emission.  Likewise, Cherenkov radiation produced by an optical pulse moving superluminally in a nonlinear medium, observed in Ref.~\citep{1984PhRvL..53.1555A},
was interpreted there as being due to the electric dipole moment induced by the time-dependent field of the pulse.
In the case of a BH, we have a truly charge-neutral ``particle'' - yet, as we demonstrate, it produces \EM\ emission.

Media with refractive index larger than unity are expected in astrophysical settings. For example, strongly magnetized plasmas, with $\om_B \geq \om_p$ ($\om_p$ is plasma frequency, and $\om_B= e B/(m_e c) $ is cyclotron frequency)  support  subluminal normal modes \citep{1986ApJ...302..138B,1999JPlPh..62...65L}.
This is the plasma regime expected, {\eg}  around various types of \NSs. 
During BH-NS mergers, the BH is then moving through magnetized media that supports  subluminal  waves.

In the present paper, 
as a simplification, we assume a non-dispersive medium, with $\epsilon  = $ constant  and trivial magnetic response, $\mu = 1$.

\section{Preliminaries}
\label{Preliminaries}

\subsection{Electromagnetic emission without an \EM\ source}

First, we  discuss the principal point -  how \EM\  emission can be produced  without an \EM\ source. 
Usually, emission is produced by a concentrated collection of accelerated  real charges. Even in the case of emission by charge-neutral particle, {\eg} electric dipole, there are two closely located separate concentration of real charges that produce the emission.

Cherenkov emission is different. It is produced by the particles of the medium - the original particle provides a disturbance that shakes the particles and makes them emit \EM\ fluctuations. In the Cherenkov regime those \EM\ fluctuations add constructively  to produce radiation, while in the normal regime the fluctuations add destructively.

Still, in the conventional Cherenkov case it is the \EM\ perturbation by {\it  real} charges that shake the particles of the medium.
The present case is different - there is no \EM\ perturber;  the disturbance is provided by the \BH\ moving through the background  \EM\ fields. In the Cherenkov regime the disturbances induce fluctuations of the fields that add constructively.

Another important difference from the conventional Cherenkov emission is that for a moving BH gravitational perturbations  propagate with the speed of light.
Conceptually, one can then imagine that the source propagates with the speed of light, but the resulting \EM\ perturbations are delayed,  creating a distributed source of \EM\ fluctuations.

The relevant references to Cherenkov emission include \cite{TammFrank,1948PhRv...74..950J,1948PhRv...74.1485J,Bolotovskii&Stolyarov,Bolotovskii1957}. In particular, Refs.  \cite{1948PhRv...74..950J,1948PhRv...74.1485J} discuss Cherenkov emission in the particle frame. This frame has an advantage that the system is stationary. 
In the conventional Cherenkov case, this leads to the fact that in the particle frame there 
are stationary electric and magnetic fields (see also \cite{1974JPlPh..12..297C}). These fields 
oscillate in space with slowly decreasing amplitudes.  Radiation in the lab frame is then achieved by Lorentz transformation.

We start with the Lagrangian density in the form
\ba &&
\cL = -\frac{1}{4} g^{\mu\nu} g^{\a\b} F_{\mu\a} F_{\nu\b} 
+ \half \kap g^{\mu\nu} F_{\mu\a} u^\a F_{\nu\b} u^\b
\nn && 
\kap = \eps - 1
\ea
where $F_{\mu\a} = \part_\mu A_\a - \part_\a A_\mu$ is the Maxwell tensor, $g_{\a\b}$ is metric tensor,  $u$ is the four-velocity of the medium normalized to $u^\mu u_\mu=-1$,  and $\eps$ is the dielectric constant of the medium.

Variation with respect to the vector potential $A_\rho$  gives
\be
\partial_\mu (\sqrt{-g} g^{\mu\nu} g^{\rho\b} F_{\nu\b} )  =   \kap \left( 
  \part_\mu ( \sqrt{-g} g^{\mu\nu} u^\rho u^\b F_{\nu\b})
-  \part_\a ( \sqrt{-g} g^{\rho\nu}  u^\a  u^\b F_{\nu\b})  \right) 
\label{1}
\ee
Eq. (\ref{1}) determines the evolution of \EM\ fields in a dielectric medium in the presence of 
a moving black hole.
It amounts to a generalization to curved space of the flat-space expressions presented in Refs. 
\cite{1948PhRv...74..950J} [Eq.~(14) with gauge Eq.~(10)] and \cite{Bolotovskii&Stolyarov} 
[Eqs.~(3.4)--(3.5)]; see also \S \ref{flat}. 
In vacuum $\kap=0$, while in the absence of gravity  $g_{\a\b} \to \eta _{\a\b}=\{-1,1,1,1\}$.

Equation (\ref{1})  has the structure
\be
\hL[g_{\mu\nu},u^\mu] A = 0 \, ,
\label{hL}
\ee
where $A$ is the electromagnetic four-potential, and $\hat{L}$ is a linear operator. 
Formally, Eq. (\ref{hL}) has no source term.
The only possible of source of radiation then is the deformation of the metric and the four-velocity 
of the medium from their flat-space values---effects start at the linear order in 
the gravitational radius  
$R_g = 2 M$.
Our goal
will be to develop a perturbation theory in this parameter and use it to compute the deformations of the 
electric and magnetic field to the first order in $R_g$.

Qualitatively, the gravity of the BH distorts the \Bf,
and these distortions are velocity dependent. For the
fields Fourier transformed with respect to the direction of propagation,
we find that in the normal regime the distortions decay exponentially with the cylindrical radius, while in the Cherenkov regime they form outgoing radiation.

 To develop a perturbation theory in $M$,  we transform to the rest frame of the BH, and decompose the operator $\hL$ in (\ref{hL}) as
\be
\hL = \hL_0 + \hL_1 \, ,
\label{decomp}
\ee
 where $\hL_0$ corresponds to  a uniformly moving dielectric in flat space, and $\hL_1$ is the remainder. We then look for a solution to (\ref{hL}) in the form
\be
A_\mu = \Azero_\mu + \Aone_\mu + \dots ,
\label{pert_exp}
\ee
where $\Azero_\mu$ is a choice of the zeroth-order approximation, which satisfies $\hL_0 \Azero = O(M)$, and $\Aone$ is the $O(M)$ deformation; the dots stand for higher-order terms. Then, correct to the first order in $M$, eq. (\ref{hL}) can be rewritten as 
\be
\hL_0 A^{(1)} = - (\hL_0 + \hL_1) \Azero \equiv S\, ,
\label{hh0}
\ee
where the source $S$ is $O(M)$.

\subsection{Waves in moving dielectric in flat space} 
\label{flat}

Equation (\ref{hh0}) has the fluctuating part of the field ``on the left" and the source part ``on the right". 
The operator $\hL_0$ acting on the fluctuating part
is none other than the wave operator in a uniformly moving dielectric, in the 
absence of any gravitational effects 
(see \S  \ref{velocityfield} and Appendix \ref{velperp} for discussion of the concept of``uniformly moving''  in the case of curved space).
Its structure has been discussed in Refs.~\cite{1948PhRv...74..950J,1948PhRv...74.1485J,Bolotovskii1957}. Here, we list a few of its properties that we will need in the following.

First, we introduce the Cherenkov parameter
\be
\Lambda^2  \equiv 1 -  \gamma^2 \beta^2 \kappa
\ee
where $\beta = v/c$, and $\gamma$ is the corresponding gamma factor.
 The parameter $\Lambda$ becomes zero at 
 \be
 \beta = \frac{1}{\sqrt{\epsilon} }  = \beta_{ph} 
 \ee
 where $ \beta_{ph}  = 1/\sqrt{\epsilon}$ is phase velocity of natural modes. Thus, 
the condition $\Lambda =0$ defines the transition from the normal to Cherenkov regime. 

Particularly important is the choice of the gauge condition. Following \cite{1948PhRv...74..950J} we adopt
\ba &&
\Pi^{\mu\nu} \part_\mu \Aone_\nu = 0 \, ,
\nn &&
\Pi^{\mu\nu} \equiv g^{\mu\nu} - \kappa u^\mu u^\nu =  
\left(
\begin{array}{cccc}
 -(1+ \gamma ^2 \kappa ) & 0 & 0 & \beta  \gamma ^2 \kappa  \\
 0 & 1 & 0 & 0 \\
 0 & 0 & 1 & 0 \\
 \beta  \gamma ^2 \kappa  & 0 & 0 & \Lambda ^2 \\
\end{array}
\right)
\label{guage}
\ea

Upon this gauge choice, the operator in question becomes
\[
[\hL_0 \Aone]^\rho = \Pi^{\mu\nu} \part_\mu \part_\nu \Psi^\rho \, ,
\]
where
\be
\Psi^\mu = \Pi^{\mu\nu} \Aone_\nu \, .
\label{Psi}
\ee
Then, eq.~(\ref{hh0}) can be rewritten as
\be
\Pi^{\mu\nu} \part_\mu \part_\nu \Psi^\rho = S^\rho \, ,
\label{fo2}
\ee
where $S^\rho \equiv - [\hL_1 \Azero]^\rho$.

Having found $\Psi$ from (\ref{fo2}), we can invert (\ref{Psi}) to obtain
\be
\Aone_\nu = \widetilde{\Pi}_{\nu\rho} \Psi^\rho
\label{inv}
\ee
where \cite{1948PhRv...74..950J}
\be
\widetilde{\Pi}_{\nu\rho} = g_{\nu\rho} + \frac{\kap}{1 + \kap} u_\nu u_\rho =
\left(
\begin{array}{cccc}
-1+  \frac{ \kappa }{1+\kappa }\gamma ^2  & 0 & 0 & \frac{ \kappa }{1+ \kappa } \beta  \gamma ^2\\
 0 & 1 & 0 & 0 \\
 0 & 0 & 1 & 0 \\
 \frac{ \kappa }{1+\kappa} \beta  \gamma ^2& 0 & 0 & 1+ \frac{ \kappa }{1+\kappa }\beta ^2 \gamma ^2  \\
\end{array}
\right)
\, .
\ee
 Alternatively, we can apply $\widetilde{\Pi}$ directly to (\ref{fo2}), to obtain an equation for $\Aone$ with  a modified source:
\[
\Pi^{\mu\nu} \part_\mu \part_\nu \Aone_\sigma = j_\sigma \, ,
\]
where
\[
j_\sigma = \widetilde{\Pi}_{\sigma\rho} S^\rho \, .
\]
Thus, the gauge choice (\ref{guage}) results in
a pair of 
separate equations for the transverse ($x$ and $y$) components of $\Aone$ and another pair for mixed $0-z$  components.

 As a simple application, consider propagation of waves in an
isotropic moving dielectric in the absence of any sources ($S=0$). In this case, Eq.~(\ref{fo2}) has plane-wave solutions, of the form $\propto e^{- i ( \om t- k ( z \cos\theta + x \sin \theta))}$, for which the equation becomes
\ba &&
{\rm Disp}  \times 
\Pi^{\mu\nu}
 \times \Aone_\nu \equiv  {\rm Disp} \times \Psi^\mu=0 
\nn && 
{\rm Disp}  = k^2 -\omega ^2-\gamma ^2 \kappa  (\omega -\beta  k \cos \theta)^2
\label{disp1}
\ea
 Eq.~(\ref{disp1}) determines the dispersion relation of the waves.
 
  The
same relation can be derived directly by the Lorentz transformation of the four-vector $\{\om; {\bf k}\}$ from the frame of the dielectric:
\ba &&
{k'}^2 = \epsilon {\om'}^2
\nn &&
k^{ 2}- \om^{ 2} = 
{k'}^2 - {\om'}^2=
( \epsilon -1) {\om'}^2 =
\kappa \gamma^2 ( \om -   k \beta \cos \theta)^2
\label{kk}
\ea
(in Eq. (\ref{kk}) values in the frame of the dielectric are denoted with primes).
In the BH frame $\om =0$ at Cherenkov resonance, so (\ref{disp1}) becomes
\be
 \cos ^2\theta =\frac{1}{\beta ^2 \gamma ^2 \kappa } =
 \frac{1}{1- \Lambda ^2} = \frac{1}{1+| \Lambda |^2} 
 \label{dd1}
 \ee
(the velocity now refers to that of the medium in the BH's frame). This is the Cherenkov cone in the BH frame (for $\Lambda ^2 < 0$). 

To recover the Cherenkov cone in the lab frame $\theta_C$ we need to use relations for the aberration of light in a medium,
\be
\cos \theta_C= \frac{  \left(\sqrt{\epsilon } \cos \theta -\beta \right)}{\sqrt{
   \left(1-\beta  \sqrt{\epsilon } \cos \theta \right)^2+(\epsilon -1)/\gamma^2}} = \frac{1}{\beta  \sqrt{\epsilon }}
   \ee

Finally, plane  waves in moving dielectric can be derived using Lorentz transformations of the \EM\ fields. This results in  relations that mix  polarization and  dispersion. It is  the power of the gauge choice (\ref{guage}) that allows simple ``unlinking'' of the dispersion and polarization, Eq. (\ref{disp1}).

\subsection{Isotropic coordinates}
Below, we calculate the source term $S$ in (\ref{hh0}) for a \BH\ moving in a magnetized dielectric. The calculation is done in the BH frame,
which has the advantage that the metric and the fields are all static.  In addition,
for the case of parallel propagation, there are
no electric fields in this frame.

Our calculations will be done in isotropic coordinates.
For motion along the \Bf, \S \ref{along},  we use
 cylindrical  isotropic coordinates $t-\rho-\phi-z$ in the $-+++$ convention. The diagonal metric terms are 
\ba &&
   g_{00} =- \frac{\left(1-\frac{M}{2 \sqrt{\rho ^2+z^2}}\right)^2}{\left(1+\frac{M}{2 \sqrt{\rho ^2+z^2}}\right)^2}
   \nn &&
   g_{\rho \rho }= g_{zz} = \left(1+ \frac{M}{2 \sqrt{\rho
   ^2+z^2}}\right)^4
   \nn &&
   g_{\phi \phi } = \left(1+ \frac{M}{2 \sqrt{\rho
   ^2+z^2}}\right)^4 \rho^2
   \label{isotr}
   \nn &&
   g\equiv \sqrt{-{\rm det} g_{\mu \nu} } =  \left(1-\frac{M}{2 \sqrt{\rho ^2+z^2}}\right) \left(1+\frac{M}{2 \sqrt{\rho ^2+z^2}}\right)^5 \rho
   \ea
   with $G$  and $c$ set to unity.  Isotropic coordinates allow most natural account of both spherical (GR) and Cartesian (linear motion of the \BH) geometries. 

   For motion across the \Bf, \S \ref{across1}, we use Cartesian isotropic coordinates.

Using expansion in weak gravity we then   solve for the  dynamics of the \EM\ perturbations to the stationary solution. In short, the \EM\ perturbations die out exponentially in the regular regime, and behave as waves (standing waves in the BH frame)  with slowly decreasing amplitudes in the Cherenkov regime.

\subsection{Choice of the velocity field}
\label{velocityfield}

 We also need to impose the velocity structure $u^\mu$. In our approach, we are not solving for the dynamics of matter. Motion of the medium is important, though: in a moving dielectric  \EM\ energy  both  propagates and is advected by the bulk motion.

Thus, there is some freedom in choosing the velocity structure. Our goal here is a demonstration of the principle, rather
than a computation for a specific astrophysical environment.

As a model example, we consider  ``straight-line'' motion along the $z$ axis, by which we mean that the radial and azimuthal cylindrical components of the velocity are both zero (note that this definition is coordinate choice-dependent).
In other words we assume that 
\be
u^\mu = \{ u^0(\x), 0 ,0 , u^z(\x) \} \, ,
\label{ugen}
\ee
where the components are some functions of the
spatial coordinates, subject to the condition
$u^2 = -1$ and such that, at large distances from
the BH, 
\be
u^\mu \to \left\{ \gamma  ,0,0,
-  \beta  \gamma \right\} 
\label{ulim}
\ee
with $\beta > 0$.
Eq.~(\ref{ulim}) represents, in the frame of the BH, an asymptotic flow of the medium in the negative $z$ direction.
For illustrative purposes, we will sometimes use
\be
u^\mu = \left\{ \frac{\gamma}{\sqrt{|g_{00}|}},0,0,
-  \frac{\beta  \gamma}{\sqrt{g_{zz}}}
\right\}  
\label{u}
\ee
where $g_{00}$ and $g_{zz}$ are the corresponding metric coefficients.

In astrophysical application, the ``straight-line'' approximation  may be justified if we are interested at distances 
larger than Bondi–Hoyle–Lyttleton radius for a moving BH.
In that case, only matter with the impact parameter smaller than the
corresponding Bondi radius $R_B = 2GM/v^2$
(for supersonic motion) is affected by the BH. At
larger radii the intrinsic kinetic pressure of the medium opposes gravity, so that for $r\geq R_B$ the hydrodynamic  reaction of the medium compensates for gravity, and the 
velocity of the medium remains mostly  unaffected by the \BH.
Since at NS-BH mergers we expect $v \sim c$, the Bondi radius is small, of the order of the \Sch\ radius.

\subsection{Stationary perturbation in BH frame}

We may expect that, when the asymptotic velocity of the 
fluid relative to the BH is small, the deformation of the fields is smooth and 
decaying away from the black hole.
The question we wish to ask is if, at a sufficiently large velocity, the nature of the 
deformation changes, so that it can no longer can be extended smoothly to the entire space.

There is then a subtlety: 
in terms of Fourier transforms with respect to $z$, that means that 
we must now choose between the outgoing or incoming asymptotic of the fields at
infinity. The outgoing one corresponds to the Cherenkov radiation. Thus, there is implicitly some time-dependence, namely, that the source was turned-on at some time in the far past. (This  relates, {\eg} to the Landau  rule \cite{1960ecm..book.....L}).

Expansion of fluctuations in terms of planar waves is not practical in our case since the waves need to be connected to  a particular effective source.

After the Fourier transform in $z$, Eq.~(\ref{fo2}) becomes
\be
\left[ \Delta _\perp -k_z^2 \Lambda^2 \right]
\left(
\begin{array}{c}
-   \left[(1+\gamma^2 \kappa) A_0  + \beta \gamma^2 \kappa A_z \right]
\\ 
A_x
\\
A_y
\\
  \left[- \beta \gamma^2 \kappa A_0 + \Lambda^2 A_z \right]
\end{array}
\right)=
\left(
\begin{array}{c}
S^0
\\ 
S^x
\\
S^y
\\
S^z
\end{array}
\right)
\label{SSS}
\ee
where $\Delta _\perp = \partial_x^2+\partial_y^2$,
 and the sources on the right are the Fourier transforms of the original.
The differential operator on the left-hand side of (\ref{SSS}) is the same for all the components---this is the power of the gauge choice (\ref{guage}).
We can therefore form convenient linear combinations of the $0$ and $z$ components.
A particularly clear choice is
\ba &&
A_1 = \beta A_z -A_0
\nn && 
A_2 = A_z-  \beta A_0
\label{A12}
\ea
for which the sources are 
\ba
\nn &&
S_1 = \frac{ \beta S^z+ S^0}{1+\kappa}
\nn &&
S_2= S^z+\beta S^0
\label{S12}
\ea
This can be inverted, as follows:
\ba &&
A_0 = \gamma^2 (A_2\beta -A_1)
\nn &&
A_z = \gamma^2(A_2-\beta A_1)
\nn &&
S^0 = \gamma^2 \left( S_1(1+\kappa)  -S_2 \beta \right) 
\nn &&
S^z= \gamma^2 \left( S_2 -(1+\kappa) \beta S_1 \right) 
\label{S13}
\ea

As a result,  all dispersion relations are  of the form
\be
 \left[
 \Delta _\perp- k_z^2 \Lambda ^2  
\right] A_i =S_i
\label{generalEq}
\ee
for $i=1,2,x,y$.

As a check, for 
 plane waves propagating at an angle $\theta$ to the
 $z$ direction,
 \be
  \left[
 \Delta _\perp- k_z^2 \Lambda ^2  
\right] \propto \sin ^2\theta  + \Lambda ^2 \cos ^2\theta 
\ee
 gives  
 a Doppler-boosted dispersion relation (\ref{dd1}).

\section{Cherenkov emission of a \BH\ propagating along \Bf}
\label{along}

Above in \S \ref{Preliminaries} we outlined two ingredients: distortion of the \EM\ fields by the gravity of a \BH\ and \EM\ processes in moving media. Next, we combine them to discuss a new effect, Cherenkov emission by a \Sch\ \BH.

We start with the mathematically simpler case of propagation along the field. In this case the toroidal coordinate is cyclic. 
Importantly, for  parallel propagation  (in the sense that the velocity of the BH is along the \Bf\ {\it at infinity}) there is one specific no-emission case: when at each point the velocity is along the local \Bf.

\subsection{Different choices of source}
\label{Newperturb}

Next we develop a perturbation theory to solve Eq. (\ref{1}) to the first order in $M$. As the earlier discussion implies, there are different version of this theory, corresponding to different choices of the zeroth-order approximation $\Azero$ in (\ref{pert_exp}). In particular, observe
that a BH causes a deformation of the fields, which starts at the first order in $R_g$, even in the absence of a medium. We may choose to include this first-order term in the unperturbed solution or we may consider it as a part of the perturbation.

 Different choices of $\Azero$ will lead to instances of Eq.~(\ref{hh0}) that differ in the structure of the source and hence lead to different solutions for $\Aone$. Upon combining the solution for $\Aone$ with the corresponding zeroth-order term, we recover the result independent of the perturbation scheme. Here, we consider two such choices, which offer independent insights.

\subsubsection{Source choice I}
\label{SourcechoiceI}
 In the first choice, we include the deformation of the magnetic field by the BH in the zero-order approximation. 
That is,
we start with 
 the  \EM\  one-form potential (flux function) in the presence of a \Sch\ \BH\
\be
 A^{(0)}_\phi =
\frac{ B_0 \rho ^2 }{2} \left(1+\frac{M}{2 \sqrt{\rho ^2+z^2}}\right)^4
\label{S00}
\ee
This potential satisfies Maxwell's equations in vacuum in the presence of the BH, 
$\partial_\mu (\sqrt{-g}  F^{\mu \nu} ) =0$.  As a result, in this case 
the source term comes only from the right hand side of (\ref{1}). 
Since  it is  explicitly proportional to $\kappa\equiv \epsilon -1$, 
this form of the source stresses the dielectric nature of the effect.

\begin{figure}[h!]
\includegraphics[width=.99\linewidth]{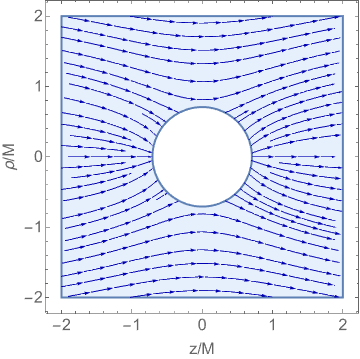}
\caption{Magnetic field lines around \BH\  in isotropic coordinates.}
\label{Bfield-iso}
\end{figure}

 The components of the magnetic field
corresponding to (\ref{S00})  and computed according to \cite[\S 90]{LLII}
are (Fig. \ref{Bfield-iso})
\ba &&
B^\rho = \frac{M \rho  z}{\left(\rho ^2+z^2\right)^{3/2}} 
\times \left(1+ \frac{M}{2 \sqrt{\rho ^2+z^2}}\right)^{-3} B_0
\approx \frac{M \rho  z}{\left(\rho ^2+z^2\right)^{3/2}} B_0 
\nn &&
B^z = \left(  1+ \frac{M \left(z^2-\rho ^2\right)}{2 \left(\rho ^2+z^2\right)^{3/2}} \right) \times  \left(1+ \frac{M}{2 \sqrt{\rho ^2+z^2}}\right)^{-3} B_0
\approx \left(1-\frac{M \left(2\rho ^2+ z^2\right)}{\left(\rho ^2+z^2\right)^{3/2}} \right) B_0
\ea

\subsubsection{Source choice II}
\label{SourcechoiceII}

As the second choice of source, which turns out to be mathematically somewhat simpler, 
we use as our unperturbed solution the uniform
magnetic  field (and, for the case of perpendicular propagation, its companion electric field arising from the boost),
corresponding to the flat space. 

Thus, in this case the unperturbed potential  for parallel propagation is
\be
 A^{(0)}_\phi =
\frac{ B_0 \rho ^2 }{2}.
\label{S000}
\ee
 As a result, for our choice of the velocity, Eq.~(\ref{ugen}), the 1-form $u^\mu \Fzero_{\mu\nu}$ vanishes, so 
the source term now comes only from the left hand side of (\ref{1}), {\it i.e.}, from the vacuum Maxwell term. 
(This is not the case for perpendicular propagation, where both sides contribute to the source.)

The main advantage of this choice is that the relations are generally simpler. The second choice also highlights an important point:  parameter $\kappa$ does not appear simply as an expansion parameter.

\subsection{Main calculations: parallel propagation}
\label{Main00}

 For parallel propagation, it is convenient to use the cylindrical version of the isotropic coordinates, in which
the metric has the form (\ref{isotr}). The source then has only the azimuthal component, with the different choices discussed
above corresponding to (see Appendix \ref{math} for a collection of relevant mathematical relations)
\be
S^\phi = M  B_0 
\left\{
\begin{array}{cc}
 \frac{ \rho  \left(2 z^2 - \rho ^2\right)}{\left(\rho ^2+z^2\right)^{5/2}} \times   \beta ^2 \gamma ^2 \kappa,    & \mbox{Choice I}
 \\
 2  \part_\rho \frac{1}{(\rho^2 + z^2)^{1/2}} , & \mbox{Choice II}
 \end{array}
 \right.
\ee

 As a consequence of
our gauge choice, this source 
will produce a nontrivial deformation of $A_\phi$ only, 
$\Aone_\phi(\rho, z)$,
which is axially symmetric, and the equation for which can be brought to the Bessel form by the substitution
\be
A^{(1)}_\phi(\rho, z)  = \rho a(\rho, z) \, .
\label{a}
\ee

The equation becomes
\be
\part_\rho^2 a + \frac{1}{\rho} \part_\rho a - \frac{a}{\rho^2} + \Lambda^2 \part_z^2 a = S^\phi \, .
\label{eqa}
\ee
This should be solved together with 
the regularity condition $A_\phi = 0$ at $\rho = 0$.

Going over to the Fourier transform with respect to $z$,
\[
\ta(\rho, k_z) = \int_{-\infty}^\infty dz e^{-ik_z z} a(\rho, z) \, ,
\]
we obtain the equation
\be
\part_\rho^2 \ta + \frac{1}{\rho} \part_\rho \ta - \frac{\ta}{\rho^2} - \Lambda^2 k^2 \ta 
= - M B_0 k 
\left\{
\begin{array}{cc}
  2 \left(1-\Lambda ^2\right) \rho  K_0\left(\rho  k\right)   & \mbox{Choice I}
 \\
 4  K_1(k \rho)
, & \mbox{Choice II}
 \end{array}
 \right. 
\label{eqta}
\ee
where $K_{0,1}$ are  the modified Bessel functions, and $k \equiv |k_z|$. It is clear that this has a particular
solution proportional to $K_1(k\rho)$; that solution, however, is singular at $\rho \to 0$. We must add
to it the solution to the corresponding homogeneous equation, chosen so as to remove the singularity.
The result is
\be
\ta(\rho, k_z) = \frac{ M B_0}{k}  \times 
\left\{
\begin{array}{cc}
\frac{4 \Lambda  K_1(\Lambda  k \rho )}{1-\Lambda ^2}
-\frac{4 K_1(k \rho )}{1-\Lambda ^2}-2 k \rho  K_0(k
   \rho ), & \mbox{Choice I}
\\
\frac{4 \Lambda  K_1(\Lambda  k \rho )}{1-\Lambda ^2} - 
\frac{4}{ 1 - \Lam^2} K_1(k \rho) 
, & \mbox{Choice II}
 \end{array}
 \right. 
\label{sol}
\ee

 Recall that, upon multiplication by $\rho$ and the inverse Fourier transform, this has to be added to the corresponding
zeroth-order term $\Azero$, to produce the full vector potential to the required order. $\Azero$, however, does not bring in any $\Lambda$ dependence. As a result, the Cherenkov terms $\propto K_1(\Lambda  k \rho )$ must coincide for the two
choices in (\ref{sol}), and we see that they indeed do.

\begin{figure}[h!]
\includegraphics[width=.99\linewidth]{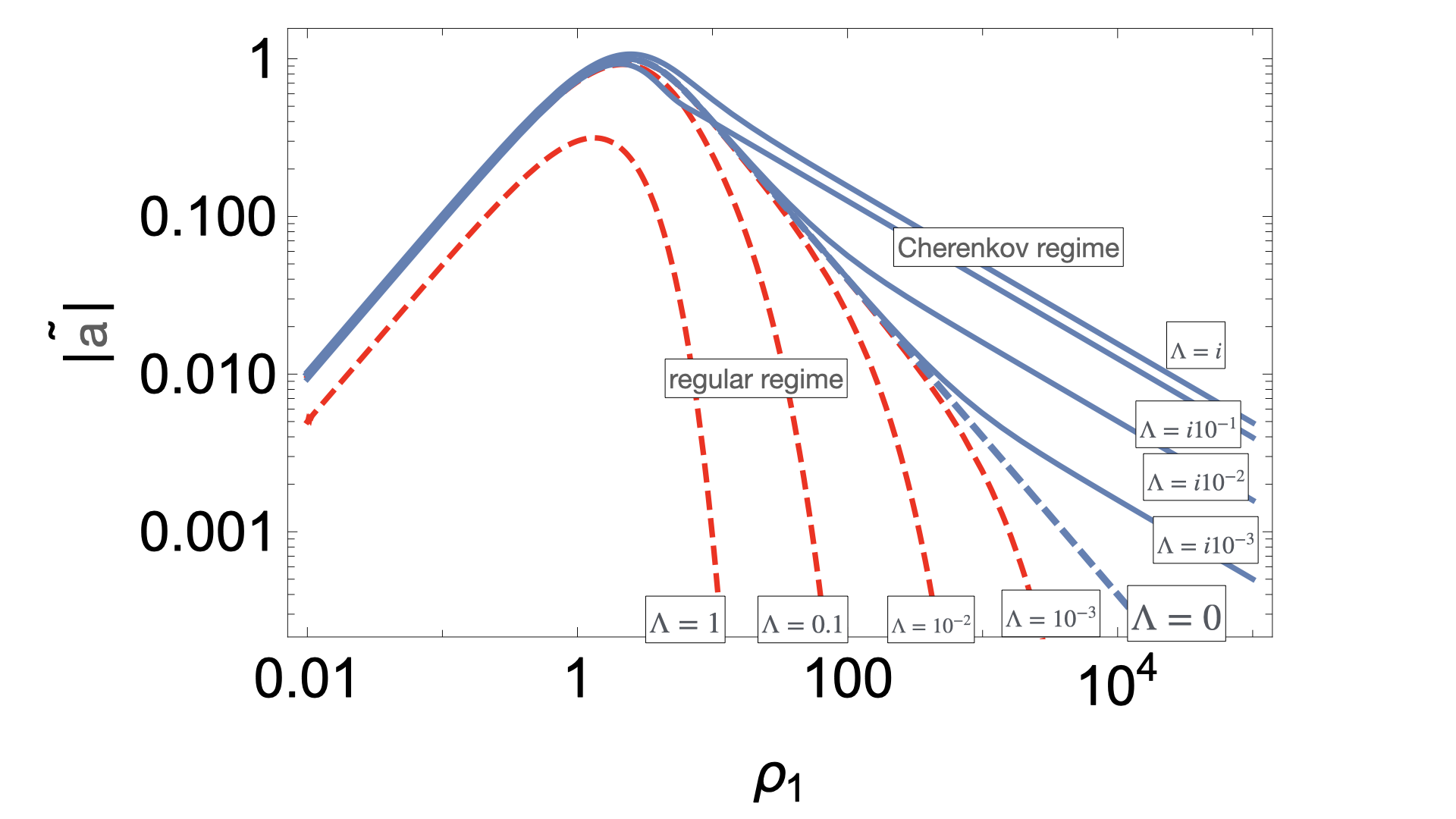}
\caption{Vector potential $|\tilde{a}|$ (excluding normalization $M B_0/k$), Eq. (\ref{sol}), Choice I.  The $\Lambda =0$ curve is given by (\ref{Lambda0}) - it clearly serves as a separatrix between regular and Cherenkov regimes. }
\label{BH-Cherenkov1}
\end{figure}

 The magnitude of (\ref{sol}) for Choice I is plotted, for different values of $\Lam$ in Fig. \ref{BH-Cherenkov1}. The    
special solution  for $\Lambda =0$, the borderline between the regular and Cherenkov regimes, is 
 \be
 \tilde{a}  \propto \frac{4}{\rho _1} -   2 \rho _1 K_2 
 \label{Lambda0}
 \ee
 For small values of $\rho_1$, this goes as $\tilde{a}  \propto \rho_1$; for $\rho_1 \gg 1$
it decays as a power-law,  $ \tilde{a} \propto 4/ \rho _1 $.

Its nature of the solution is different in the normal regime (real $\Lambda$) and the Cherenkov regime (complex $\Lambda$).
 We focus on the term involving $K_1(\Lam k \rho)$. In the normal regime,  
$\Lambda ^2 > 0$,  
 at large distances, this term is fast decreasing:
     \be
 \tilde{a} \propto   
 \frac{ e^{ -  \Lambda k \rho } }{( k \rho ) ^{1/2} }
    \label{24} 
     \ee
In contrast, in the Cherenkov regime,  $\Lambda ^2 < 0$, 
   \be
   \tilde{a} \propto  \frac{e^{\pm i |\Lambda |  k \rho}}{\sqrt{k \rho}}
    \label{Solution1}
   \ee
(We will return to the choice of the sign in this expression shortly.)

The solution (\ref{Solution1}) shows an oscillatory behavior, with decreasing amplitude $\propto \rho ^{-1/2}$, characteristic of
the cylindrical waves of Cherenkov radiation.

It is also instructive to go over to the coordinate space. 
We will do that starting with Eq.~(\ref{sol}) for Choice II. 
 In the non-Cherenkov regime, $\Lam^2 > 0$, the first-order deformation $\Aone_\phi(\rho, z)$ is 
obtained directly by multiplying (\ref{sol}) by $\rho$ and Fourier transforming. The result is
\be
\Aone_\phi(\rho, z) = \frac{2 M B_0}{1 - \Lam^2} \left[ (\rho^2 + z^2)^{1/2} - (\Lam^2 \rho^2 + z^2)^{1/2}
\right] \, .
\label{solA}
\ee

Observe that in the absence of the medium, {\it i.e.}, for
$\Lambda \to 1$, 
Eq. (\ref{solA})
becomes
\be
\Aone_\phi(\rho, z) =
\frac{M B_0 \rho^2}{(\rho^2 + z^2)^{1/2}} \, .
\label{exp}
\ee
This precisely reproduces the $O(M)$ term in the expansion of the exact solution (\ref{S00}).  Thus, the same term that is included as a part of the zeroth-order
approximation for Choice I, appears as a part of the first-order solution for Choice II.

 Observe also that, close to the boundary with the Cherenkov regime, $\Lam^2 \to 0$, the second term in the bracket in (\ref{solA}) approaches $|z|$ (at a given
finite $\rho$), meaning that the magnetic field undergoes a rapid charge near the $z=0$ plane---a precursor to the formation of the Cherenkov leading shock-like wavefront.

 In the Cherenkov regime, $\Lam^2 < 0$, 
the term in $\Aone_\phi(\rho,k_z)$ responsible for Cherenkov radiation is
\be
A_\phi^{(rad)}(\rho, k_z) =\frac{4 \Lambda  }{1 - \Lambda ^2} \times  \frac{ M B_0}{k} \rho   K_1( \Lambda  k \rho ) \, .
\label{Arad}
\ee
Going over to the coordinate space 
requires analytically continuing this to
$\Lambda = \pm i|\Lambda|$  (using the procedure described in Subsec.~\ref{Emittedpower} below) and Fourier transforming the result. 
We find 
\be
A_\phi^{(rad)}(\rho, z) = - \frac{2 M B_0}{1 + |\Lam|^2} 
\left\{ \begin{array}{lr} z , & |z| < |\Lam| \rho \\
z + (1 - \mbox{sgn\,} z) \sqrt{z^2 - |\Lam|^2 \rho^2}  , & 
|z| \geq |\Lam| \rho
\end{array} \right. 
\ee
where $\mbox{sgn\,} z$ denotes the sign of $z$.
This potential is smooth for $z > 0$ ({\it i.e.}, in front of the black hole) but 
exhibits a shock-like wavefront at $z=-|\Lam| \rho$. The direction normal to this wavefront coincides with that of Cherenkov emission, Eq. (\ref{dd1}).

\subsection{Emitted power}
\label{Emittedpower}

Given the vector potential in the Cherenkov regime (\ref{Arad}) one can find the \Bf\ in the BH frame,
 using the expressions $B_z = (1/\rho) \partial_\rho A_\phi$ and 
$B_\rho = -(i k_z / \rho) A_\phi$.
After the boost along $z$ to the lab frame,  the radial component $B_\rho$ gives rise
to a tangential component of the electric field $E_t' = - \gamma v B_\rho$.
In addition, we need to replace $z$ with $z = \gamma (z' - vt')$ in the formulas of the inverse
Fourier transform as, for instance, in
\be
E_t' = - \gamma v \int_{-\infty}^{\infty} \frac{dk_z}{2\pi} e^{ik_z \gam (z' - v t')} B_\rho \, .
\label{Et}
\ee
The total power radiated in the Cherenkov regime through a cylindrical surface at a large $\rho$ is given by 
\be
{\cal P} = \frac{1}{4\pi} 2\pi \rho \int_{-\infty}^\infty dz' E_t' B_z'
= - \frac{1}{2} v \rho \int_{-\infty}^\infty \frac{dk_z}{2\pi} B_\rho B_z^* \, .
\label{power}
\ee
(Note the disappearance of $\gamma$ in the last expression.) Here, it is sufficient to use
the slowly decaying radiation parts of the fields, obtained by an analytical continuation to imaginary $\Lambda$. Eq.~(\ref{Et}) shows that this continuation should be carried out differently, depending on the sign of $k_z$. Indeed, for $k_z > 0$, the frequency 
\[
\ome' = \gam k_z v 
\]
in (\ref{Et}) is positive, so that the outgoing cylindrical wave is given by the Hankel function
$H^{(1)}(|\Lam| k \rho)$, 
which is obtained via
\[
\Lam = - i |\Lam| \, .
\]
For $k_z < 0$, the frequency is negative, and the outgoing wave is $H^{(2)}(|\Lam| k \rho)$, corresponding to $\Lam = i |\Lam|$. As a result, the Fourier transforms of the radiation 
fields required by (\ref{power}) are
\ba 
B_\rho & = & - \frac{4 M B_0 |\Lam|}{1 + |\Lam|^2}  K_1(\mp i |\Lam| k \rho) \, , 
\nn 
B_z & = & \frac{4 M B_0 |\Lam|^2}{1 + |\Lam|^2} K_0 (\mp i |\Lam| k \rho) \, ,
\label{BB}
\ea
where the upper sign corresponds to $k_z > 0$, and the lower to $k_z < 0$.

 Substituting (\ref{BB}) 
in (\ref{power}) and using the large-argument asymptotics of the Bessel functions, we obtain
\be
{\cal P}_\parallel = 2 \beta \left( \frac{M B_0 |\Lam|}{1 + |\Lam|^2} \right)^2 \int_{-\infty}^{\infty} \frac{dk_z}{k} 
\label{Ppara}
\ee

 Estimating maximal $k_z \sim 1/R_g$ and minimal  $k_z \sim 1/R_s$, the  size of the system, we obtain the total luminosity as
 \be
 {\cal P} _\parallel\approx \frac{\beta |\Lambda| ^2}{\left(1+|\Lambda |^2\right)^2} \times 
 ( B_0   R_g) ^2 c 
 \times \log \left(\frac{R_s}{R_g}\right)
 \label{LCh} 
 \ee
where we reinstated dimensional quantities. 

\section{Motion across \Bf}
\label{across1}

\subsection{The source}
Next we consider the case of BH moving across \Bf. 
 This case turns out to be somewhat  mathematically more complicated than the motion along the \Bf,  since there is no cyclic coordinate. 

\begin{figure}[h!]
 \includegraphics[width=.99\linewidth]{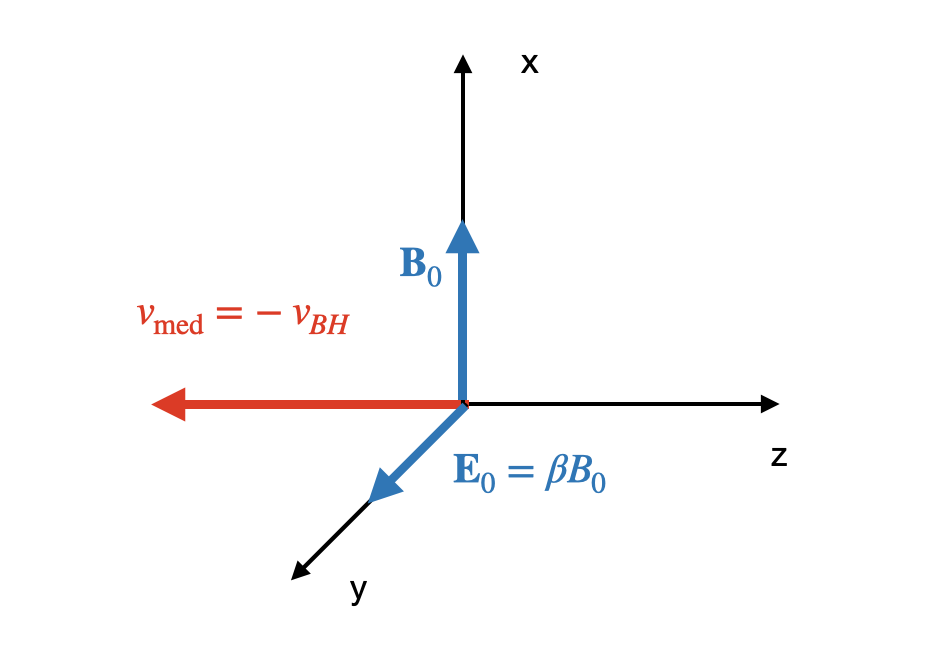}
\caption{Choice of coordinates for \BH\ moving perpendicular to \Bf. In the frame of \BH\ the medium is moving with velocity $-v_{\rm BH} $  (along $-z$),  producing   crossed magnetic and electric fields at infinity. }
\label{E0B0-perp}
\end{figure}

In the \BH\ frame there are now crossed electric and magnetic fields at infinity. 
We chose the velocity of the \BH\ along $z$ and the asymptotic magnetic field along $x$, so that the asymptotic 
electric field is along $y$ (Fig.~\ref{E0B0-perp}).
The isotropic Cartesian coordinates all have the same diagonal elements, equal  to $g_{zz}$ in Eq.  (\ref{isotr}). 

Similarly to the case of parallel propagation, we have freedom in choosing the zeroth-order state. The {\it Choice-I}   is to start with gravity-distorted fields, so that
in the \BH\ frame, the four-potential for constant \EM\ fields in the symmetric gauge  is  (note new notation for the fields)
\ba &&
A = \left\{ 1,0,0,0\right\}  \left(1-\frac{M}{2 r}\right)^2 y \cE +
\nn &&
+ \left\{0,0,
- z/2, y/2\right\}  \left(1+\frac{M}{2 r }\right)^4 \cB 
\nn &&
\cE = \beta \gamma B_0
\nn &&
 \cB =\gamma B_0
\nn &&
r= \sqrt{\rho ^2+z^2}
\nn &&
\rho ^2= x^2+y^2
\label{Choice1}
\ea 
where $B_0$ is the \Bf\ in the lab frame, and $\cE$, $\cB$ are the asymptotic values of the electric and magnetic fields in the BH frame.

Alternatively, we can treat the distortion as a part of the fluctuating field. In this  {\it Choice-II} the unperturbed state is 
\be
A = \left\{ y \beta,0,- z/2, y/2\right\} \cB \, .
\label{Choice2pepre}
\ee

We have already demonstrated that for parallel propagation the two choices result in the same radiative term. As it turns out, computations for the choice (\ref{Choice2pepre}) are somewhat  simpler, so we proceed with it here.

 In this section, we choose the 4-velocity of the medium in the form (\ref{u}). The corresponding expressions for a more general (one-parameter) choice are given in 
Appendix \ref{velperp}.

We find the first-order sources for equation ({\ref{SSS}}) (in the $t,x,y,z$ coordinates) as
\be
S^\rho = \left( 
\begin{array}{c}
 -\frac{2 \beta  \rho  \left(1+\gamma ^2 \kappa \right) \sin \phi }{\left(\rho
   ^2+z^2\right)^{3/2}} \\
 0 \\
 \frac{2 \Lambda ^2 z}{\left(\rho ^2+z^2\right)^{3/2}} \\
 -\frac{2 \Lambda ^2 \rho  \sin \phi }{\left(\rho ^2+z^2\right)^{3/2}} \\
\end{array}
\right) 
\times \cB M
\ee
where $\cB = \gamma B_0$ is the \Bf\ in the BH frame,
 and $\phi$ is the azimuthal angle. (Component $S_x=0$ is due to interference of E-produced and B-produced terms, each is non-zero independently.) 

The Fourier transform in $z$ gives
\be
S^\rho = \left( 
\begin{array}{c}
 -4 \beta  \left(1+ \gamma ^2\kappa \right) k  K_1\left( \rho k \right) \sin \phi  \\
 0 \\
 -4 i \Lambda ^2 k_z K_0\left(\rho  k\right) \\
 -4 \Lambda ^2 k  K_1\left(\rho  k\right) \sin \phi \\
\end{array}
\right) 
\times \cB M
\label{SSF}
\ee
The $0$ and $z$ components of this can be combined 
according  to (\ref{S12}),
\be
\left( 
\begin{array}{c}
 S_1 \\
 S_2 \\
\end{array}
\right)= - 
\left(
\begin{array}{c}
\frac{  2+\kappa }{1+\kappa }  \beta \\
1+\beta^2 \\
\end{array}
\right)
\times 4 \cB M k  K_1\left(\rho  k\right) \sin \phi
\ee
for use in the general equation (\ref{generalEq}).

\subsection{Fluctuating components and Poynting power for perpendicular propagation}

Calculations of emission for perpendicular case mostly follow the procedures described in \S \ref{Main00} and \S \ref{Emittedpower}. There are a few new effects, though.

For perpendicular propagation the source separates into 
 three parts : $S_y$, $S_0$ and $S_z$. Each component will produce correspondingly three fluctuating potentials $A_y$, $A_0$ and $A_z$.   In the BH frame the corresponding \EM\ fields are stationary. Boosting to the lab frame, we then obtain propagating waves. Importantly, even though each fluctuating component can be calculated independently, the resulting Poynting  flux includes also cross-terms.

The y-component of the general equation (\ref{generalEq})  with the source from (\ref{SSF}) has a solution 
\be
A_y = - \frac{4 i}{k_z} \left[ K_0\left(\rho _1\right)-K_0\left(\Lambda  \rho
   _1\right)\right] \times \frac{ \Lambda ^2 }{1-\Lambda ^2} \times \cB M
   \label{Aykz}
   \ee
 where $\rho_1 = k\rho$.

    The  radiative part of (\ref{Aykz}) is
   \be
   A_y^{(rad)} =\frac{4 i \Lambda ^2 K_0\left(\Lambda  \rho _1\right)}{k_z \left(1-\Lambda ^2\right)}\times \cB M 
   \label{Ay1}
   \ee

For the $A_0-A_z$ components,
using sources (\ref{SSF}) and fluctuating fields combined according to (\ref{S12}), we
 find
\ba &&
A_1= 
\frac{\beta  (2+\kappa )}{(1+\kappa) \left(1-\Lambda ^2\right) } 
\left[ \Lambda 
   K_1\left(\Lambda  \rho_1 \right) - K_1\left(\rho_1 \right)\right]
 \times  \frac{4  \cB M \sin \phi }{ k } 
   \nn && 
   A_2= 
   \frac{1+\beta ^2}{1-\Lambda ^2}
   \left[ K_1\left(\rho
   _1\right)-\Lambda  K_1\left(\Lambda  \rho _1\right)\right]\times
   \frac{4  \cB M \sin \phi }{ k} 
   \ea
These can be combined to form  $A_{0,z}$ according to (\ref{S12}),
 \ba &&
  A_0 = - \frac{\beta}{1+\kappa}   \frac{\Lambda^2}{1-\Lambda^2} \left[ \Lambda 
   K_1\left(\Lambda  \rho_1 \right) - K_1\left(\rho_1 \right)\right]
 \times  \frac{4  \cB M \sin \phi }{ k } 
 \nn &&
 A_z =  \frac{1+\gamma ^2 \kappa}{(1+\kappa) \left(1-\Lambda ^2\right) }\left[ \Lambda 
   K_1\left(\Lambda  \rho_1 \right) - K_1\left(\rho_1 \right)\right]
 \times  \frac{4  \cB M \sin \phi }{ k } 
 \ea
 It may be verified that the vector $\{ A_0,0, A_y,  A_z\}$ satisfies the gauge condition (\ref{guage}).

We can write for the radiative part $\propto K_{0,1} \left(\Lambda  \rho  k_z\right)$
   \ba && 
   A^{(rad)} = \left\{c_0 K_1\left(x\right)  \sin (\phi ) ,0, \frac{ik_z c_y}{k} K_0\left(x\right),c_z K_1\left(x\right)  \sin (\phi ) \right\}
  \times \frac{ M \cB}{k} 
  \nn &&
 x = \Lam k \rho
\nn && 
c_0 =-\frac{4 \beta  \Lambda ^3}{(1+\kappa) \left(1-\Lambda
   ^2\right)}
   \nn &&
   c_y= \frac{4 \Lambda ^2}{1-\Lambda ^2} 
   \nn &&
   c_z =\frac{4 \Lambda 
   \left(\gamma ^2 \kappa +1\right)}{(1+\kappa) \left(1-\Lambda ^2\right)}
   \ea
Gauge condition  ensures that 
   \be
   c_0=\frac{\beta  \Lambda  \left(c_y - \Lambda  c_z\right)}{1-\Lambda ^2}
   \label{gauge1}
   \ee

 At large $\rho$, we can use the asymptotic form of the Bessel functions
   \be
   K_n (x)  \to \sqrt{\pi/(2x) }  e^{-x} \equiv f(x)
   \ee
The radiative part of the 4-potential is then
\be 
 A^{(rad)} = \{ c_0 \sin \phi,0, \frac{ik_z c_y}{k}, c_z \sin \phi \} \frac{  f(x)}{k}
\ee   

The electric and magnetic fields in the BH frame,
in the cylindrical coordinates $(\rho, \phi, z)$, become
   \ba && 
\E = \left\{ -\Lam, 0, \frac{i k_z}{k} \right\} \times
c_0 \sin \phi  \times f(x)\times M \cB  
   \nn &&
   \B = 
  \left\{c_y \cos (\phi ),\frac{c_0 \left(\Lambda^2 -1 \right) \sin (\phi )}{\beta  \Lambda
   }, - \frac{i k_z \Lam c_y}{k} \cos (\phi )\right\}      \times f(x)\times M \cB 
  \ea

 To obtain the power radiated in the Cherenkov regime, we follow the same procedure as in the case of parallel propagation. Namely, we boost these expressions to the lab frame and analytically continue the results via $\Lam \to \pm i |\Lam|$, with the sign depending again on the sign of $k_z$.

For the radial Poynting vector we find
   \be
   { F}_\rho  =  \left(c_y^2 \cos ^2(\phi ) + \frac{|c_0|^2 (1+ \kappa) \sin ^2(\phi )}{|\Lambda|^2} \right)  \times \frac{\beta  \gamma }{8 \rho  k} \times (\gamma M B_0)^2 
   \ee
   where we eliminated $c_z$ according to (\ref{gauge1}) to demonstrate that the Poynting flux at infinity is explicitly positive. 
   
   Finally the Cherenkov power for perpendicular propagation reduces to a relatively compact expression 
   \ba && 
   {\cal P}_\perp = 2 \beta \times   \left( \frac{|\Lambda| ^2}{1+|\Lambda| ^2} \right)^2  \times 
   \left(\cos ^2(\phi ) + \frac{\beta ^2 \sin ^2(\phi )}{\epsilon }\right)
 \times  \frac{dk_z}{2  \pi k}  \times (\gamma M B_0)^2
 \nn &&
  {\cal P}_{\perp,tot}= \int {\cal P}_\perp d \phi =
 \beta  \left(1+ \beta^2/\epsilon\right)  \left( \frac{|\Lambda| ^2}{1+|\Lambda| ^2} \right)^2\times  \frac{dk_z}{k}  \times (\gamma M B_0)^2
 \label{Pperp}
   \ea
(compare with the parallel case (\ref{Ppara})).

We may also perform inverse Fourier transforms to find fields in coordinate space
\ba && 
A_y(z, \rho) = \frac{ \Lambda ^2 }{1-\Lambda ^2} 
\ln \left(\frac{\Lambda  \left(\sqrt{\rho ^2+z^2}+z\right)}{\sqrt{\Lambda ^2 \rho
   ^2+z^2}+z}\right)
\times 2 \gam M B_0
\nn &&
A_0(z,\rho) =  \beta  \frac{\Lambda ^2}{ \epsilon \left(1- \Lambda ^2\right) } 
\left(\sqrt{\rho ^2+z^2}-\sqrt{\Lambda ^2 \rho ^2+z^2} \right)
\times \frac{ 2 \gam M B_0 \sin \phi}{\rho}
\nn &&
A_z= -\frac{1+ \gamma ^2 \kappa}{\beta  \Lambda ^2} A_0
   \label{Azrho1}
   \ea
   (Cherenkov condition is not assumed in (\ref{Azrho1}).)

 \section{Observability}

A simple order-of-magnitude estimate of Cherenkov power $L$ is given by Eqns.  (\ref{LCh}) and (\ref{Pperp}). Returning to dimensional quantities
\ba &&
L \sim \beta_K B(r)^2 \left( \frac{2 G M_{BH}}{c^2} \right)^2 c \Lambda _c 
\nn &&
 B(r)\approx B_{NS} \left(\frac{r}{R_{NS}} \right)^{-3}
\nn &&
\beta_K \sim \sqrt{ \frac{ G M_{tot} }{r c^2}}
\nn &&
\Lambda _c =\ln \frac{k_{max}}{k_{min}}
\label{LL}
\ea 
($ B(r)$ is the magnetic field at distance $r$ expressed in term of the surface \Bf, $\beta_K$ is the beta-parameter for Keplerian motion,  $ M_{tot}$ is the total mass of the system, and $\Lambda _c$ is  a parameter similar to Coulomb logarithm). 

Consider first a merger of a stellar mass  BH   $M_{BH} \geq$ few $\times M_\odot \gg M_{NS}$  (so that $M_{tot} \sim M_{BH}$) merging with a \NS\ of surface \Bf\ $B_{NS}$.

Using time to merger in post-Newtonian approximation  \cite{1994PhRvD..49.2658C}
\be
-t= \frac{5}{256} \frac{c^5}{G^3} \frac{r^4}{M_1 M_2 M_{tot}},
\ee
(estimate of $-t$ is valid approximately to $-t \sim$ milliseconds), 
we find 
 \be
L_{\rm BH, 10 M_\odot} \approx 4 B_{NS}^2  \frac{(G M_{BH})^{2} (G M_{tot})^{1/2} R_{{NS}}^6}{c^4 r^{13/2}} \Lambda _c   \approx 10^{38} \, {\rm erg\, s}^{-1} (-t)^{-13/8} 
\left( \frac{B_{NS}}{10^{12} {\rm G}}  \right) ^{2}\left( \frac{M_{BH}}{10 M_\odot}  \right) ^{-3/4}
\label{LCH0}
 \ee
 where $-t$ is time to merger in seconds,   we assumed that the total mass is dominated by the BH, and approximated $\Lambda_c \sim \ln  R_{NS}/(2 G M_{BH} /c^2) \approx 1$. The peak power, estimated at $-t \sim $ millisecond is $\sim $ few $10^{42}$ erg s$^{-1}$.   (Incidentally, the value of peak power comes close to the parameters of enigmatic Fast Radio Bursts \cite{2019ARA&A..57..417C}, but the expected corresponding rates naturally do not match.)
  
  On the one hand, the  peak luminosity is not high, similar to other mechanisms  based on unipolar inductor  scaling \citep{2001MNRAS.322..695H,2011PhRvD..83l4035L}. But it can be detectable in radio - the corresponding peak flux is in the kilo Janskys range if 
 coming from $\sim 100$ Mpc distance \cite{2024arXiv240216504L}.

   An important constraint comes from the expansion in small mass of the BH, which translates to the requirement that the emitted wavelength should be larger than the \Sch\ radius. For stellar mass BHs the  \Sch\ radius is kilometers, this implies frequencies below the kilo-Hertz range. Depending on the parameters at the source, such low frequencies may  not be able to propagate  due to plasma cut-off at $\omega = \omega_p $.  The required  density is $\leq 1$ cm$^{-3}$. This is somewhat larger  than the typical density in the volume-dominant warm phase of ISM ($n _{ISM} \sim 10^{-2}- 10^{-1} $ cm$^{-3}$). 
   
   Primordial black holes \citep{2021RPPh...84k6902C} may  not suffer from such limitation -  their emitted frequency can be in the GHz range for masses $M_{BH} \sim 10^{-6} M_\odot$.  The corresponding power for a primordial black hole flying through the \ms\ of a neutron star
  can be estimated using (\ref{LL}) with $B\sim B_{NS}$ and    $\Lambda_c \sim \ln  R_{NS}/(2 G M_{BH} /c^2) \approx 15$. 
  We find 
  \be
  L_{\rm BH, 10^{-6} M_\odot, max} \approx 4 
  \times 10^{34} \, {\rm erg\, s}^{-1}
\label{LCH1}
 \ee

    \section{Discussion}
We discuss a novel effect at the intersection of general relativity, electromagnetism, and the physics of continuous media: Cherenkov emission by an uncharged superluminally moving \Sch \BH. Interestingly, in this case there are no source terms: emission originates as a perturbation-of-a-perturbation: the initial \EM fields are distorted by gravity, and in the Cherenkov regime the perturbations of those distorted fields propagate as \EM\ waves.

In the Cherenkov regime,  emission conditions involve the combination
\be
|\Lambda|  B_0 M \equiv 
\sqrt{ |\beta^2 \gamma^2 \kappa-1|} B_0 (G M )
\label{factor}
\ee
(see  Appendix \S \ref{velperp} for an interesting exception).

Scaling (\ref{factor}) indicates that the effect is the interplay between gravity $(G M )$, magnetic field $B_0$, and physics of movement of continuous media: both $\kappa \neq 0$, $\beta \neq 0$ are needed, plus the Cherenkov condition.

    The process involves a number of subtle physical effects:
    \begin{itemize}
        \item 
   In Cherenkov emission, it is the medium that emits. Conventionally, a moving particle produces an \EM\ perturbation - these perturbations force the particles of the medium to emit coherently in the Cherenkov regime.  In our case, it is the gravity of the \BH\ that ``shakes'' the particles of the medium.
   \item Gravity propagates with the speed of light, but the corresponding disturbances \EM\ are delayed. This leads to a fairly complicated effective source structure.
   \item Important simplification is achieved by transforming to the \BH\ frame. Our approach for treating fluctuations follows the conventional Cherenkov case. In this frame, all fields are stationary.
    \item  The emitted spectrum is red-dominated $\propto d k_z/|k_z|$. Formally, the lowest emitted frequency is then determined by the size of the system, but for low frequencies, our assumptions will be violated ({\eg} plasma cut-off effects).
    \end{itemize}

An important free ingredient in the model is the choice of the velocity field,  as we do not solve for the motion of matter.
 The results in general depend on that choice. In particular,
there is one special case when there is no Cherenkov emission. It occurs for 
  a \BH\ propagating along the direction of the \Bf\  at infinity in the regime when {\it locally}  the matter flows along the \Bf.

Note also that
one cannot gain much in a highly dielectric medium with $\epsilon \gg 1$ (possibly large values of $\Lambda$) since Cherenkov
powers (\ref{LCh}) and (\ref{Pperp}) typically level off as functions of $\Lambda$.

We thank Roger Blandford and Ilya Mandel for discussions and comments on the manuscript. 

\bibliographystyle{apsrev}
  \bibliography{./BibTex}

\appendix 
\section{Relevant mathematical formulae}
\label{math}
In many cases we deal with the equation
\ba && 
\Delta_\perp f- \Lambda^2 f=S
\nn &&
S= Z(\rho_1) e^{i m  \phi}
\ea
with  $m=0, \, 1$; $Z$ is a Bessel function of orders $0$ or $1$, $\rho_1$ is dimensionless cylindrical coordinate.

The homogeneous equation, $S=0$,
\be
\Delta_\perp f- \Lambda^2 f= 0
\ee
has solutions 
\ba &&
f= K_0(\Lambda  {\rho_1})
\nn &&
f= K_1(\Lambda  {\rho_1}) e^{i  \phi}
\ea

The following  are the relevant solutions of the 
inhomogeneous equation
(note the ones for 
$S = \rho_1 Z(\rho_1) e^{im\phi}$, which do not seem to be listed in the standard textbooks on Bessel functions)
\ba && 
S= K_0(\rho_1) \to f= \frac{1}{1-\Lambda^2} K_0(\rho_1)
\nn &&
S= \rho_1 K_1(\rho_1) \to f = \left(\rho _1 K_1\left(\rho _1\right)+  \frac{2 K_0\left(\rho _1\right)}{1-\Lambda ^2}\right) \frac{1}{1-\Lambda^2} 
\nn &&
S= K_1(\rho_1) e^{i  \phi} \to f=  \frac{1}{1-\Lambda^2} K_1(\rho_1) e^{i  \phi}
\nn &&
S= \rho_1 K_0(\rho_1)e^{i  \phi} \to f=
 \left({\rho _1 K_0\left(\rho _1\right)}+ \frac{2 K_1\left(\rho _1\right)}{\left(1-\Lambda
   ^2\right)} \right)
   \frac{e^{i \phi }}{1-\Lambda ^2}
   \ea

   The following combinations of the homogeneous and inhomogeneous solutions are well-behaving at $\rho_1 \to 0$
\ba && 
S= K_0(\rho_1) \to f= \frac{1}{1-\Lambda^2} \left( K_0(\rho_1) - K_0(\Lambda \rho_1) \right)
\nn &&
S= \rho_1 K_1(\rho_1) \to f= \left(\frac{2 K_0\left(\rho _1\right)}{1-\Lambda ^2}-\frac{2 K_0\left(\Lambda  \rho
   _1\right)}{1-\Lambda ^2}+\rho _1 K_1\left(\rho _1\right) \right)  \frac{1}{1-\Lambda^2} 
\nn &&
S= K_1(\rho_1) e^{i  \phi} \to f= 
\left( K_1(\rho_1) - \Lambda K_1(\Lambda \rho_1)  \right) \frac{e^{i \phi }}{1-\Lambda ^2}
\nn &&
S= \rho_1 K_0(\rho_1)e^{i  \phi} \to f=
 \left(\rho _1 K_0\left(\rho _1\right)+\frac{2 \left(K_1\left(\rho _1\right)-\Lambda  K_1\left(\Lambda  \rho
   _1\right)\right)}{1-\Lambda ^2} \right)
   \frac{e^{i \phi }}{1-\Lambda ^2}
   \ea
   The radiation components correspond to terms dependent on  $(\Lambda  \rho_1)$.

The importance of these relations is: perturbation theory may start with different assumptions about the zeroth-order state. For example, we may chose flat metric and treat effects of gravity and of the medium as perturbations, or we may chose vacuum solutions in curved space, and add matter effects. {\it  Different choices of the zeroth-order state result in different sources $S$, but in the same radiation components.}

\section{Choices of velocity field for perpendicular propagation}
\label{velperp}

For perpendicular propagation, 
unlike the case of motion parallel to the field, there is no longer a universal result for arbitrary ``straight-line" motion. In other words, the result depends on the 
form of the coordinate dependence in
\[
u^{\mu} = [ u^0({\bf x}), 0, 0, u^z({\bf x}) ]
\]
Here, we consider expressions for the 4-velocity that are of the form
\[
u^{\mu} = [ \gam (1 + a/r), 0, 0, - \gam \beta (1 + b/r) ]
\]
where the constants $a$ and $b$ are $O(M)$, chosen so that the condition 
$u^2 = -1$ is satisfied to the first order in $M$.  The condition $u^2 = -1$ results in
the relation
\[
M - a + \beta^2 ( M + b) = 0
\]
To the {\em zeroth} order, i.e., for
$u^\mu_{(0)} = (\gam, 0, 0, -\gam \beta)$ in the BH frame, the condition $u^\mu_{(0)} \Fzero_{\mu\nu}$ holds:
\[
\gam \Fzero_{0y} - \gam \beta \Fzero_{zy} = 0
\]
where
\[
\Fzero_{0y} = - \gam \beta B_0 \, , \hspace{3em} \Fzero_{zy} = - \gam B_0 \, ,
\]
are obtained as a result of the Lorentz transformation of the magnetic field to the BH frame.
To the first order, however, 
\[
u^\mu \Fzero_{\mu y} \approx (b - a) \gam^2 \beta M B_0 \frac{1}{r} = - 2 C M \beta B_0 \frac{1}{r}
\]
where we have defined
\[
C \equiv \frac{1}{2 M} \gam^2 (a - b)
\]
Unless $a=b$, this is nonzero, meaning that the right-hand side of (\ref{1}) now contributes to 
the source. The model used in the main text corresponds to
$a = -b = M$, and $C = \gam^2$.

The components of the source $S^\a \equiv - [L_1 \Azero]^\a$ 
in (\ref{fo2}) are now obtained, in parallel with those  for the parallel motion
as
\ba
S^0 & = & - \part_y (\sqrt{-g} g^{yy} g^{00} \Fzero_{y0}) - \kap \gam \part_y (u^\mu \Fzero_{\mu y})
\approx 2 M \gam v B_0 (1 + C \kap) \part_y \frac{1}{r} \, , \\
S^x & = & 0 \, , \\
S^y & = & - \part_z (\sqrt{-g} g^{zz} g^{yy} \Fzero_{zy}) - \kap \gam \beta \part_z (u^\mu \Fzero_{\mu y})
\approx - 2 M \gam B_0  (1 - C \kap \beta^2) \part_z \frac{1}{r} \, , \\
S^z & = & - \part_y (\sqrt{-g} g^{yy} g^{zz} \Fzero_{yz}) + \kap \gam \beta \part_y (u^\mu \Fzero_{\mu y})
\approx 2 M \gam  B_0 (1 - C \kap \beta^2) \part_y \frac{1}{r} \, ,
\ea
where we have used the components of the metric correct to the first order in $M$.

Recall that these are the sources for $\Psi^\mu$, which is related to $\Aone_\mu$ by Eq.~(\ref{inv}):
\[
\Aone_\nu = \widetilde{\Pi}_{\nu\rho} \Psi^\rho \, .
\]
The sources for $\Aone_\mu$ then are
\[
j_\mu = \widetilde{\Pi}_{\mu\nu} S^\nu
\]
for which we find
\[
j_0 = C_0 \part_y \frac{1}{r} , \hspace{3em}
j_x = 0, \hspace{3em}
j_y = C_y \part_z \frac{1}{r} , \hspace{3em}
j_z = C_z \part_y \frac{1}{r}
\]
where 
\ba
C_0 & = & 2 M \gam B_0 \beta \left[ - (1 + C\kap) + \frac{2\kap \gam^2}{1 + \kap} + \frac{C \kap^2}{1 + \kap} \right] \\
C_y & = & - 2 M \gam B_0  (1 - C \kap \beta^2) \\
C_z & = & 2 M \gam B_0 \left[ 1 - C\kap \beta^2 + \frac{2\kap \gam^2 \beta^2}{1 + \kap} + \frac{C \kap^2 \beta^2}{1 + \kap} \right]
\ea
 There is a relation between the coefficients,
\be
C_y + \kap \gam^2 \beta C_0 + \Lam^2 C_z = 0 \, ,
\label{Crel}
\ee
which is a consequence of the gauge condition (\ref{guage}). Using this relation, we can bring the total power radiated in the Cherenkov regime through a cylindrical surface at a large $\rho$ to the form
\be
{\cal P} = \frac{\beta}{4 (1 + |\Lam|^2)^2}\left( C_y^2 + \eps C_0^2 \right)
\int_{-\infty}^\infty \frac{dk_z}{k} \, ,
\label{Cpower}
\ee
where as before it is understood that the integral needs to be cut off both at small and at large wavenumbers.

 A curious aspect of the formula (\ref{Cpower}) is that it shows that using a different model for the velocity than the one in the main text in general changes the threshold ($\Lam \to 0$) behavior. Indeed, for the model in the text,
$C_y$ and $C_0$ are both proportional to $|\Lam|^2$, which results in an overall factor of $|\Lam|^4$ in the power.
We see, however, that this not the generic behavior: generically, the power sets in at a nonzero value immediately once the Cherenkov regime is reached. This is the behavior argued previously to occur for the Cherenkov radiation by a magnetic dipole moving superluminally in a medium and oriented perpendicular to the 
direction of motion \citep{1984SvPhU..27..772F}

\end{document}